# Highly efficient and broadband wide-angle Holography Using Patch-Dipole Nano-antenna Reflectarrays

Yuval Yifat, Michal Eitan, Zeev Iluz, Yael Hanein, Amir Boag, and Jacob Scheuer*

School of Electrical Engineering, Tel Aviv University, Tel Aviv 69978, Israel

**ABSTRACT:** We demonstrate wide-angle, broadband and efficient reflection holography by utilizing coupled dipole-patch nano-antenna cells to impose an arbitrary phase profile on of the reflected light. High fidelity images were projected at angles of 45° and 20° with respect to the impinging light with efficiencies ranging between 40%-50% over an optical bandwidth exceeding 180nm. Excellent agreement with the theoretical predictions was found at a wide spectral range. The demonstration of such reflectarrays opens new avenues towards expanding the limits of large angle holography.

Computer generated holography is a widely used technology in diverse applications, ranging from authentication and optical data storage, to interferometry, particle trapping and phase conjugation [1-4]. Holograms typically utilize dielectric structures realized by laser writing [5], direct machining or e-beam lithography [6]. These methods facilitate the realization of holographic elements which radiate complex waveforms efficiently to small angles or simple grating lobes to large angles through specific conditions of resonant gratings [7-9]. Wide angle projection of computer generated holograms requires a steep phase gradient between adjacent pixels. Attaining such gradient necessitates a small number of pixels in each period and renders the projected hologram inefficient [10-12]. Thus, efficient projection of complex waveforms to large angles still remains a challenge [13, 14].

A highly attractive alternative approach for generating complex reflective patterns is utilizing nano-antenna elements. Nano-antennas are nano-meter sized metallic structures which scale down the concept of conventional Radio-Frequency (RF) antennas and resonate at optical frequencies [15-18]. Such components have been the subject of continuous research during the past decade owing to their unique capabilities for light manipulation and attractiveness for applications such as energy harvesting [19-21], non-linear optics [22-24] and sensing [25-27]. Similar to their lower frequencies counterparts, nano-antennas enable control over the phase and amplitude of the light scattered from them and, owing to their small size, can generate phase discontinuities in length scales much smaller than the wavelength of light, as described by the generalized reflection and refraction laws [28-30]. Using contemporary fabrication and design capabilities, nano-antennas can serve as attractive building blocks for diverse optical functionalities such as lensing [31,32] beam deflection [33] and complex beam-shape shaping [34,35]. Recently, efficient reflection methods have been introduced (about 14%-27% in [32] and 80% in [33]), suggesting that reflection arrays are highly suitable for beam control applications.

Recent studies have demonstrated the use of nano-antennas for holography [36-38], opening up a new approach for forming holographic structures. However, so far mostly optical transmission modes were utilized, limiting the measured efficiency below 10%. As a new alternative we propose and demonstrate here the use of a nano-antenna reflectarray for efficient, broadband and wide angle holography applications. Very recently, there has been a report [39] of using nano-antennas over a reflecting surface to create polarization dependent holograms, but the reported efficiency was less than 20% overall and about 10% for reflection of the hologram to an angle of 45°.

In this letter we experimentally demonstrate, the employment of nano-antennas for generating a broadband, highly efficient holographic image projected at large angles. Our choice of pattern for this demonstration is the logo of Tel-Aviv University (TAU) projected to angles of 20° and 45° relative to the incident beam. The efficiencies we present here exceed 40% over an operational bandwidth of at least 180 nm, which is substantially broader than previously reported results. The results show significantly higher efficiencies than that of previous reports on holograms at wide angles [10-12, 39], and approach the efficiency values of the simple waveforms projected to high angles by volume holograms [14].

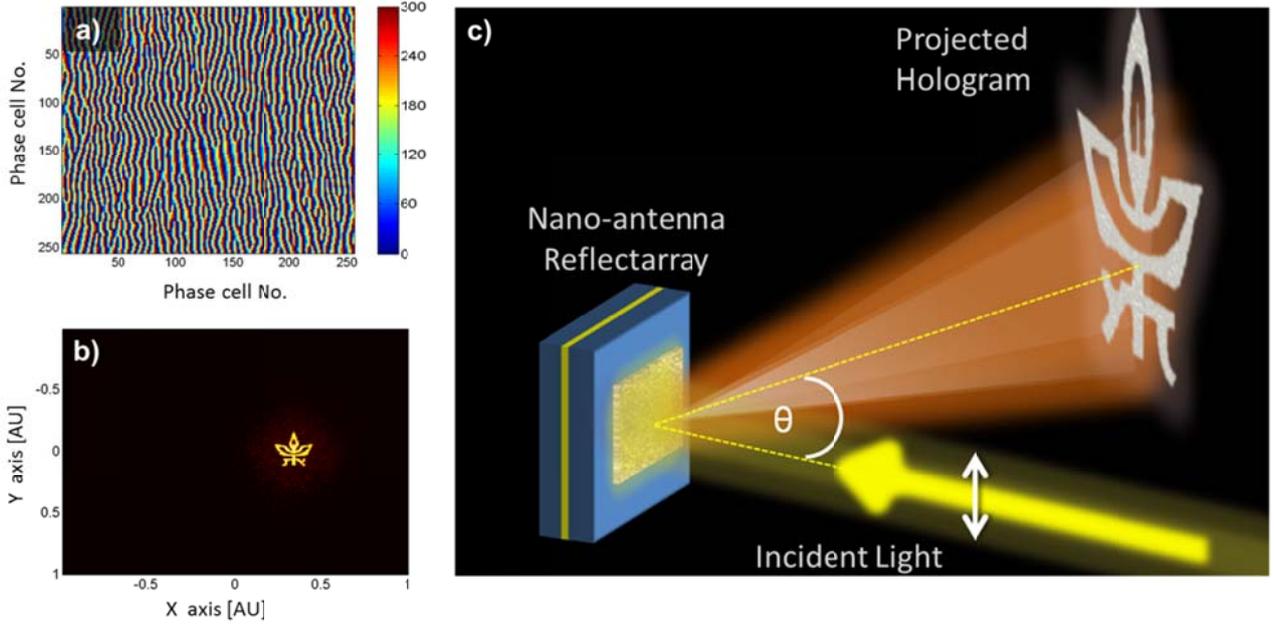

Figure 1. Phase map and resulting far-field images obtained from simulation. (a) Determined phase map with a quantization of 6 angles (color online). (b) Simulated far field image for TAU logo projected to 20° relative to incident beam. (c) Illustration of experimental concept – Linearly polarized light beam is directed towards the nano-antenna array phase map and creates a TAU logo projected to an angle θ in the far-field.

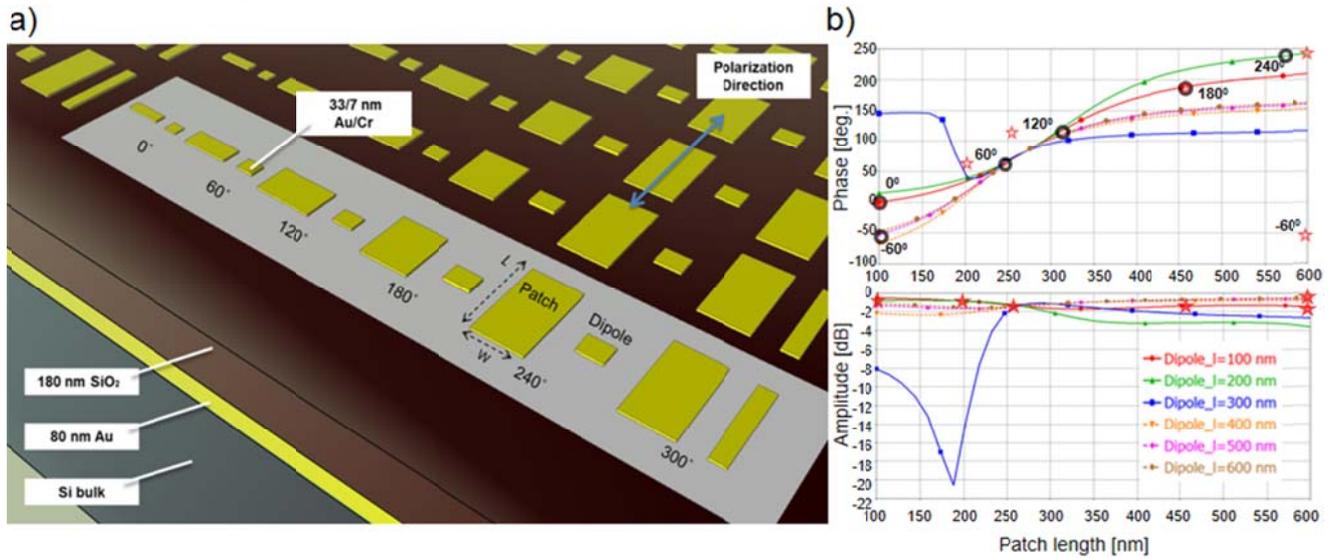

Figure. 2. Unit cell geometry: (a) top view of the patch and dipole nano-antennas unit cells and the phase response they create (left to right 0° – 300°). The exact dimensions of each element are given in Table 1; (b) phase response (top) and amplitude response (bottom) of the different antenna elements obtained from numerical simulations for individual elements in an array of identical neighbors. Circles - initial elements dimensions, Stars – elements dimensions following super-cell optimization. The loss of the individual elements for the final design is presented by the Star markings in the amplitude graph.

The realization of a nano-antenna based hologram requires, first, to determine the phase map corresponding to the desired output beam. This can be done by implementing the Gerchberg-Saxton algorithm in accordance with the work described in [40,41]. This algorithm, which is widely used for holography applications and phase retrieval problems, is provided with the (complex) amplitudes of the incident optical field and the desired far-field pattern. It then determines the required phase map at the reflection (antenna) plane through an iterative process of direct and inverse Fourier transforms. In our experiments, the antenna array is placed at the waist of a Gaussian beam with a FWHM of 500 μm with a wavelength of λ=1550 nm (similar to setup described in [42]). Note, that the incident beam is larger than the array in order to obtain illumination with constant phase and nearly uniform amplitude.

The input beam for the Gerchberg-Saxton algorithm is a constant phase Gaussian beam with the appropriate size and the output beam is the TAU logo projected to different angles. For computational reasons, the algorithm is implemented using a 256 x 256 element matrix. The algorithm is run until the normalized absolute error, i.e. the absolute difference between the desired pattern and the calculated one normalized by the pattern size, drops below a predetermined value. At this point the phase profile needed for generating the desired beam is found.

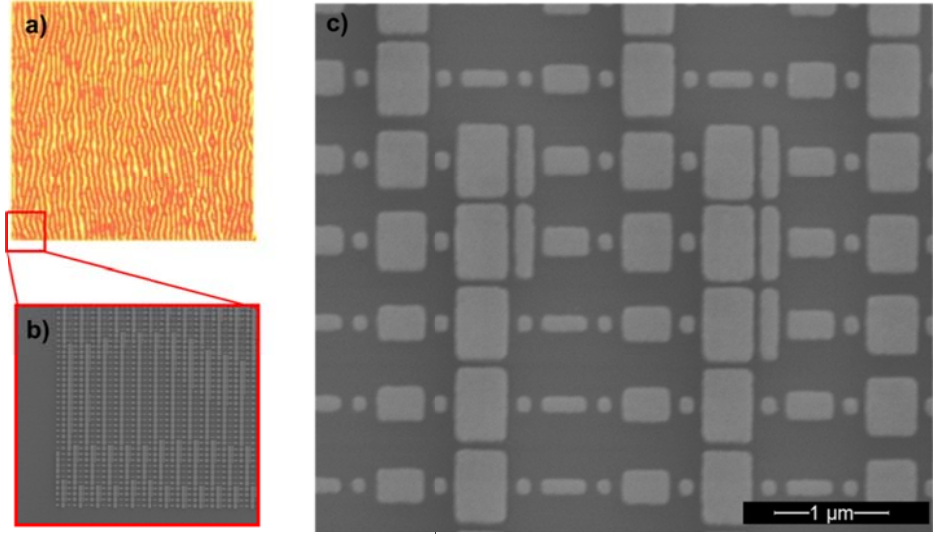

Figure 3. (a) Optical microscope image of fabricated array. (b) SEM image of corner of array. (c) High magnification image of fabricated antennas.

Fig. 1 depicts the required phase map and the corresponding optical output, as well as an illustration of the holography concept.

Following the determination of the phase map we proceed to design nano-antennas that scatter light with the desired phase. The antennas are chosen to quantize the continuous phase into the 6 discrete values between 0° and 300° with 60° increments. The quantization decreases the simulated hologram efficiency (i.e. the percentage of incident power diverted to the hologram) by approximately 15% but without noticeable change in the resulting shape. Our nano-antenna design is based on an expansion of classical antenna theory [43], taking into account the metal properties at optical frequencies [44]. In order to completely span the phase it is advantageous to utilize unit cells comprising two antenna elements of different geometries. The combined spectral response of the two elements provides more degrees of freedoms for the design and facilitates the $2\pi$ phase spanning. The choice of coupled patch-dipole antennas as the building blocks for the phase component stems from their simple fabrication and repeatability. It is quite likely that a choice of two different building blocks (e.g. a bowtie and a circular patch antenna) can also provide the necessary phase coverage and yield efficiencies which are comparable to those of the current design.

To achieve high efficiency, it is desired that the reflectivities of the individual elements of the reflectarray differ only in *phase* while maintaining constant amplitude. By properly selecting the dimensions of our antennas elements it is possible to attain a scattered wave possessing any desired phase response while retaining uniform amplitude (see Fig. 2b). Varying the dimensions of these elements alters the combined antenna resonance, which in turn changes the phase of the reflected wave.

In order to achieve proper control over the phase of the reflected wave, we simulated the coupled patch-dipole antenna structures using CST microwave studio software [45]. The layout of the different antenna dimensions is shown in Figure 2. The antennas consist of an Au layer with a thickness of 33 nm over a 7 nm Cr adhesion promotion layer. In order to increase the efficiency of the array, it is fabricated over a 180 nm SiO2 layer deposited over a reflective Au layer with a thickness of 80 nm, serving as a backplane mirror. The Au is deposited over a SiO2 substrate, which has no effect on the reflectarray response because the reflective layer is several times thicker than the skin depth of light in gold [44]. The composition of the different layers and their depths were also optimized using the CST simulation tools. The initial design of the dipole and the patch antennas in each element is carried out by analyzing their spectral response in an infinite array of identical elements. The results of the simulations of the individual elements are presented in Figure 2b. These design curves provide initial phase values from which it is possible to choose specific dipole/patch combinations. The circles in Fig. 2b indicate the choice of dipole and patch dimensions which are used as the initial design of the hologram basic building blocks, based on the design curves plotted in the figure.

Each unit cell in the hologram is 720 nm by 720 nm and consists of a dipole and a patch element. As each unit cell is an independent building block, it essentially acts as a phase pixel, enabling us to design and fabricate an array of arbitrary phase distribution. The dimensions of the unit cell are smaller than the wavelength of the incident light, resulting in strong coupling between the nano-antennas (both within a unit cell and between adjacent unit cells), which affects their plasmonic response [46] and as a result – the scattering efficiency. Consequently, the actual phase response of each element may differ substantially than that of the initial design, resulting in lower hologram efficiency. To overcome this problem, the modeling in the CST simulation is carried out by simulating a "super-cell" of the 6 phase pixels organized in sequence from 0° to 300° as shown in Figure 2a. The supercell is simulated in an infinite 2D array, which enables a computationally efficient way to optimize the elements. Modification of the elements dimensions in the supe-rcell allows for optimization of the phase response, yielding the final element dimensions detailed in Table 1. The stars in Fig. 2b indicate the dimensions of the dipole and patch nanoantennas which were obtained by the super-cell optimization. Clearly, some of the phase element dimensions differ substantially from their initial values, thus indicating the impact of coupling effects and the necessity to take them into account in order to attain enhanced efficiency. The importance of the super-cell optimization approach is further discussed in the supporting material where we present the substantial deviation of the phase responses of the original designs from the desired ones when the nanoantennas are used in a real hologram. Note, however, that even the super-cell optimization approach does not provide the optimal element design and hologram efficiency. This point is discussed in depth in the followings and in the supporting information.

| Phase | Patch Element | | Dipole Element | |
|---|---|---|---|---|
| | Height [nm] | Width [nm] | Height [nm] | Width [nm] |
| 0° | 100 | 380 | 100 | 100 |
| 60° | 200 | 380 | 100 | 100 |
| 120° | 260 | 380 | 100 | 100 |
| 180° | 460 | 380 | 100 | 100 |
| 240° | 600 | 380 | 100 | 100 |
| 300° | 600 | 380 | 600 | 100 |

**Table 1. Dimensions of elements for different phase unit cell**

The antenna arrays were fabricated using E-beam lithography in a method similar to that described in [47]. The antennas' structure is identical to that used for the numerical simulation described above. The fabricated array consist a 256 x 256 unit cells where each unit cell is a square with a side of 720 nm, yielding a final device which is 184 μm by 184 μm. Figure 3 depicts optical and scanning electron microscope images of the fabricated array.

Figure 1c illustrates the measurement setup. The incident light is emitted from a tunable laser source (λ=1450 - 1640 nm) passes through an attenuator and a polarization controller, and is collimated into free space and directed towards the sample. The sample is placed on an X-Y stage which is located at the waist of the free-space Gaussian beam. Consequently, the phase of the field impinging on the reflectarray is uniform. The impinging light beam is linearly polarized in the vertical direction, as is shown in Fig 2a.

In order to quantify the efficiency of the arrays, we measure the reflected optical power using a large area detector with an integrating sphere. The detector is first placed in front of the antennas to measure the incident optical power - $P_{in}$, and then at the point set at the center of the hologram in the far field – $P_{det}$. The efficiency is defined as:

$$\eta = \frac{P_{det}}{P_{in}} \cdot g \qquad (1),$$

where g is the geometric normalization factor arising from the size mismatch between the incident Gaussian beam (FWHM of 500 μm) and the arrays (square with a side of 184 μm). g is determined by measuring the profile of the incident Gaussian spot and calculating the actual part of the power which impinges upon the reflectarray. For the 180 x 180 μm array, this factor is approximately 17.5%. The measured wavelength dependent efficiencies of reflectarrays designed for 20° and 45° deflection are presented in Figure 4 along with an IR image of the of the TAU logo projected by the hologram.

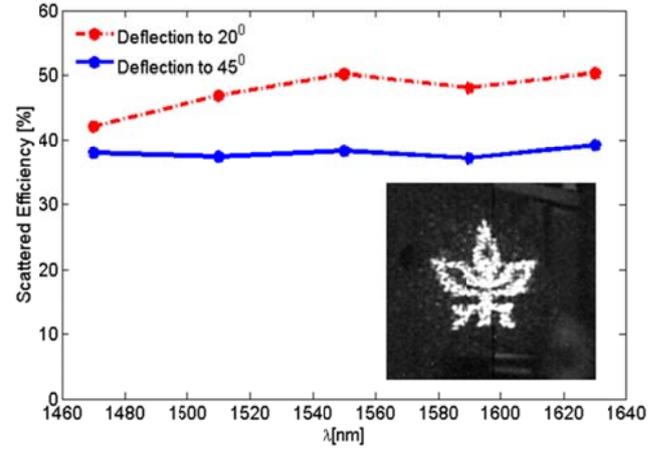

Figure 4. Efficiency measurements for TAU logo deflected to 20° (dashed red – color online) and to 45° (solid blue). Inset – Image of the logo projected to 20° taken with an InGaAs IR camera.

As shown in Fig. 4, the scattering efficiency of the hologram remains high over spectral range of 200 nm. This broadband response is due to the phase response of the designed antennas which is highly wavelength independent. The image projected by the hologram also remains unchanged (see supporting information for images and further discussion). The theoretical efficiencies of the phase quantized hologram projected to 20° and 45° are 60% and 55% respectively. The measured efficiency is slightly lower than the theoretical prediction due to fabrication errors and optimization tolerances as described further below. As may be expected, the response of the hologram depends strongly on the polarization of the impinging light. This is because of the asymmetric structure of the unit-cell [48] and the optimization procedure for a linearly polarized parallel to the antenna long axis (as indicated in Fig 2 a). When the hologram is illuminated by light which is polarized perpendicular to the antenna, the hologram efficiency decreases dramatically to the point of disappearance of the image.

In addition to the TAU logo projected to a positive angle α, and additional, image, weaker and transposed, of the logo which is scattered to a negative angle –α (see experimental and numerical results in Figures 5a and 5b respectively).

The scattering efficiency of this ghost symbol is approximately 12%. This ghost image is a well-known artifact that is pre-sent in holographic elements due to errors in their phase distribution [49,50]. The ghost is symmetric relative to the direction of incidence and corresponds to the Bragg angles of the reflectarray.

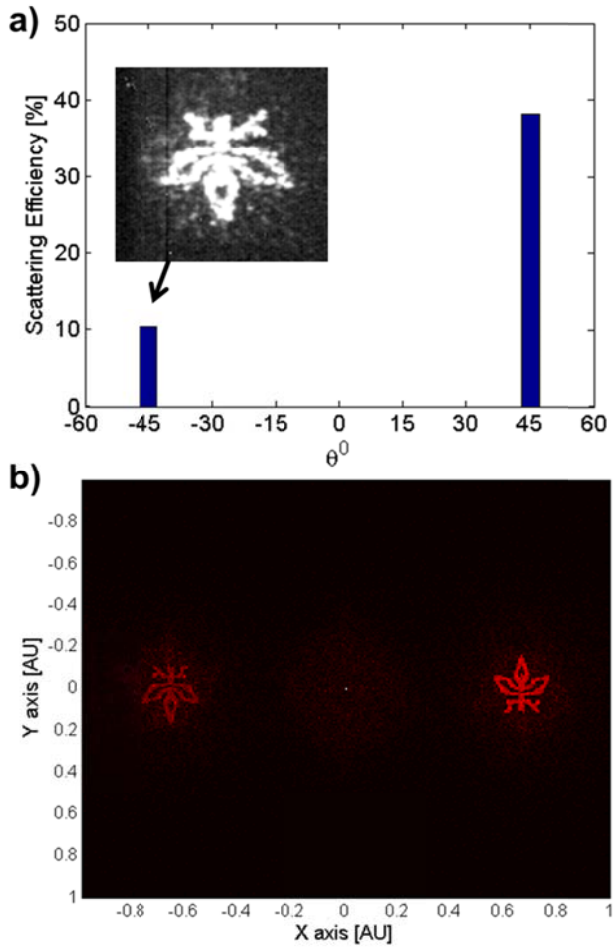

Figure 5. Efficiency measurements for TAU logo deflected to 45° and the phantom image deflected to -45°. Inset - IR camera image for inverted phantom image. (b) Simulation results for projection of original and phantom symbols. Result was obtained after incorporation of phase errors. Note appearance of zero order pattern in evident as white dot in the middle of the figure.

The phase errors in our design stem from several reasons: (a) A non-optimal solution obtained from the GS algorithm which could be improved by using advanced algorithmic techniques such as those described in [35]. (b) A loss of efficiency can be expected due to the quantization of the phase to 6 elements. Designing additional antennas will reduce phase error resulting in increased hologram efficiency and removal of the ghost image. The effect of additional phase levels was described in [51]. (c) Errors in the dimensions of some of the fabricated antenna elements due to the proximity effect [52]. (d) Coupling effects between neighboring unit cells leading to non-optimal dimensions of some of the individual elements. As mentioned earlier, the dimensions of the individual elements are determined by simulating a supercell of 6 adjacent elements corresponding to the phase responses increasing from 0° to 300°. The supercell is assumed to be part of an infinite two-dimensional array of identical supercells, and the dimensions of the individual elements in each supercell are optimized to maximize the power reflected to the desired angle. However, in the designed phase map, phase elements are often adjacent to neighbors which differ from those for which they were simulated. Thus, when a phase element is positioned between two non-consecutive elements (for example a 120° element placed between a 0° element and a 300° element), its phase response is modified with respect to the design because of the different coupling to its neighbors. In order to overcome this phase error, it is possible to improve the design by generating a look-up table which calculates the phase of each permutation of neighboring elements. Such look-up table enables the tailoring the dimensions of the individual elements to obtain the desired phase from the complete array. A deeper analysis of these phase errors is given in the supporting information.

In conclusion, we have demonstrated wide angle, highly efficient, optical holography by utilizing a reflectarray of optical nano-antennas elements comprising a coupled dipole-patch configuration to control the phase of the scattered light. We employed the GS algorithm to determine the phase map required to project the TAU logo to angles of 20° and 45° relative to the surface normal. The efficiencies of the projected hologram were found to be 40%-50% for a broad wavelength range. We observed inverted images which were projected in the opposite direction. These ghost images stem from errors in the in the realization of the phase map caused by imperfections in fabrication, phase quantization, and non-optimal antenna design. Further improvement of the hologram efficiency requires the development of methods for eliminating these phase errors. Their high efficiency over a wide bandwidth and simple fabrication flow render nano-antenna devices highly attractive for numerous applications such as beam shaping, polarization control, security or detection. Moreover, incorporating active tuning mechanism such as those described in [53, 54] it is possible to extend our approach towards the realization active holographic displays and communication devices.

## AUTHOR INFORMATION

**Corresponding Author**

\* kobys@eng.tau,ac.il

**Notes**

The authors declare no competing financial interests

## ACKNOWLEDGMENT

**The authors wish to thank Israeli DoD and Ministry of trade for partially supporting this research.**